\documentclass[12pt]{article}
\usepackage{graphicx}

\begin{document}
\begin{titlepage}
%
%






{\ }









\begin{center}
\centerline{Pressure Excitation and Ionisation:a Simple One-Dimensional Example}
\end{center}
\begin{center}
\centerline{V.Celebonovic}			
\end{center}

\begin{center}
\centerline{Institute of Physics,University of Belgrade,Pregrevica 118,11080 Zemun-Belgrade,Serbia}
\centerline{vladan@ipb.ac.rs}
\end{center}





\begin{abstract}
In interiors of celestial objects matter is subdued to extremely high values of pressure.Theoretical analysis of the behaviour of atms and molecules under high pressure is a complex quantum mechanical and statistical problem. The aim of the present letter is to demonstrate, on a known one-dimensional quantum mechanical system, the reality of excitation and ultimately ionisation of atoms and/or molecules under the influence of high external pressure.
\footnote{published in Phys.Low-Dim.Struct.,{\bf7/8},pp.127-132 (2001)}
\end{abstract}
\end{titlepage}


{

\newpage

\section{Introduction}

It is an accepted fact in physics that the only correct way of studying phenomena it atoms and molecules is the use of quantum mechanics. This branch of physics has started developing in the third decade of the last century. Solving a quantum-mechanical problem amounts to solving the Schr$\ddot{o}$dinger equation,and thus determining the energy spectrum and the wave functions. 

The first attempt of using quantum-mechanics in studying the effects of high external pressure on atomic structure is mentioned in [1],as being due to Enrico Fermi. He just "`played"' with the problem, so no definite results emerged, but he concluded that external pressure induces changes in the properties of atoms and molecules. In modern times,a quantum mechanical theory of the influence of external pressure on the atoms and molecues in a real solid was developed for the explanation of the pressure shift of the spectral lines of ruby [2]. For a recent review of quantum effects in solids underhigh pressure see, for example,[3].

This letter contains two more sections. The next one is devoted to the determination of the influence of the external pressure on the energy levels of a particle in a one-dimensional potential well, while the final part contains the conclusions.


\section{Calculations}

A one-dimensional finite potential well is defined by [4]
\begin{equation}
	V(x) = 0, |x|<a ; V_{0} , |x|>a 
\end{equation}

where $a$ denotes the half-width of the well. Let us introduce
\begin{equation}
	\alpha^{2} = 2 m E/\hbar^{2} ; \beta^{2} = 2m(V_{0}-E)/\hbar^{2}
\end{equation}
If
\begin{equation}
\xi = \alpha a ; \eta = \beta a
\end{equation}
it can be shown that the energy levels are given by [4]
\begin{equation}
\xi\tan \xi = \eta
\end{equation}

\begin{equation}
\xi^{2}+\eta^{2} = 2mV_{0}a^{2}/\hbar^{2}
\end{equation}
Introducing
\begin{equation}
	a^{2} = \frac{(n \hbar)^{2}}{2 m V_{0}}
\end{equation}
where $n > 0$ and applying it to eq.(5), one obtains 
\begin{equation}
	\xi tan \xi = \sqrt(n^{2}-\xi^{2})
\end{equation}
The energies are finally given by
\begin{equation}
	E = (\frac{\xi}{n})^{2} V_{0}
\end{equation}

Applying this algorithm for $1 < n < 10$ gives a set of values of the particle energies as a function of the width of the potential well.They can be fitted by a function of the form
\begin{equation}
	\frac{E}{V_{0}} = \sum_{i=0}^{i=5}\frac{c_{i}}{n^{i}}
\end{equation}
A~ direct calculation using one of the known data-fitting programs, gives the following calues of the constants in eq.(9) : $c_{0} = - 0.000618 ; c_{1} = 0.018006 ; c_{2} = 2.259278 ; c_{3} = - 3.678692 ; c_{4} = 2.908830 ; c_{5} = - 0.960535$ and the standard deviation of the fit is $\sigma=\pm2.2\times 10^{-6}$. Pressure is defined in statistical physics as the negative volume derivative of the energy of the system. As the system under consideration here is one dimensional, the volumeis replaced by the width of the well, which means that
\begin{equation}
	P = - \frac{\partial E}{\partial a}
\end{equation}
Application of this definition to eq.(9), leads to the following expression for the pressure:
\begin{equation}
P = \sum_{i = 0}^{i=5}\frac{i c_{i} K^{i}}{a^{i+1}}
\end{equation}
Note a change of variables in eq.(11) and 
\begin{equation}
	K = \frac{\hbar}{\sqrt{2 m V_{0}}}
\end{equation}
The pressure derivative of the energy is given by
\begin{equation}
	\frac{\partial E}{\partial P} = \frac{\partial E}{\partial a} (\frac{\partial P}{\partial a})^{- 1}
\end{equation}
After some algebra,one gets
\begin{equation}
\frac{\partial E}{\partial P}= \frac{a}{2} \frac{a^{4}c_{1}+2Ka^{3}c_{2}+3a^{2}c_{3}K^{2}+4ac_{4}K^{3}+5c_{5}K^{4}}{2a^{4}c_{1}+3a^{3}c_{2}K+6a^{2}c_{3}K^{2}+10ac_{4}K^{3}+15c_{5}K^{4}}
\end{equation}
This expression can be simplified by developing it into series in small parameters. In the case of a narrow potential well(that is, when $a\rightarrow0$), eq.(14) can be developed up second order in $a$ as a small parameter
\begin{equation}
	\frac{\partial E}{\partial P}\cong\frac{a}{6} + \frac{1}{45}\frac{c_{4}}{c_{5}}\frac{a^{2}}{K}
\end{equation}
Taking $K$ as a small parameter, physically corresponds to a deep potential well or to a massive particle. Developing up to second order in $K$, one gets the following expession
\begin{equation}
\frac{\partial E}{\partial P}\cong \frac{a}{2}- K\frac{c_{2}}{2c_{1}}+\frac{3 K^{2}}{2ac_{1}^{2}} (c_{2}^{2}-c_{3}^{2})	
\end{equation}

\section{Discussion and conclusions}

Equation (14) gives the first pressure derivative of the energy of a particle in a $1D$ finite potential well. It is the fnal result of the calculations presented in this letter. The possibility for pressure excitation (and ultimately ionisation) of this system depends on the sign ofeq.(14). This is, in turn, dependent on signs of both the denominator and the numerator, because of the fact that some of the constants $c_{i}$ have different signs. A simple algebraic analysis of eq.(15) shows that $\partial E/\partial P = 0$ for $a=a_{0}= 2.476601 K$. The physical consequences of these results are clear: when a particle in a $1D$ finite potential well is subdued to external pressure, the energy will increase and (ultimately) the system will be ionised if $a<a_{0}$. In the opposite case, the particle will be pushed deeper into the well. What about the numerical values? 

The example we are conssidering in this letter is one-dimensional and accordingly highly idealizes. However, insering into it realistic values of the atomic parameters gives useful illustrations. It is well known that the most abundant element in the universe is atomic hydrogen. Taking for the depth of the potential well $V_{0}=13.6058 eV$ and for the mass of the particle the bare electron mass, yields $a_{0} = 1.31056\times 10^{-10}m$. Since for the hydrogen atom $a=0.529\times 10^{-10}m$, it is clear that $a<a_{0}$. This implies that a $1D$ analogy of a hydrogen atom will be excited and ultimately ionised by application of sufficiently high external pressure. Pursuing the quantum mechanical aspect of the problem, one could determine the dependence of the wave function of a massive particle enclosed in a $1D$ box on the width of the box. This calculation should start from the wave function [4] 
	\[u(x)=2Ccosh[\beta x]
\]
and the value of the constant $C$ follows from the normalization condition
	\[C = \frac{1}{2\sqrt{a}}\times(1+\frac{sinh[2a\beta]}{2a\beta})^{-1/2}
\]
where $\beta$ was defined in eq.(2). Once the normalization constant thus becomes known, the probability of finding the particle within a particular interval of length within the box, as well as its changes with pressure, can in principle be determined. Square integrating the wave function $u(x)$ in the interval $-\gamma a \leq x \leq \gamma a$ (with $\gamma > 0$),one gets the following expression for the probability of finding a particle in the interval $-\gamma a \leq x \leq \gamma a$
\begin{equation}
	R = \frac{2a\beta\gamma + \sinh[2a\beta\gamma]}{2a\beta+\sinh[2a\beta]}
\end{equation}
From eqs.(2) and (9) one obtains
	\[\beta^{2} = (\frac{2m}{\hbar^{2}})\times V_{0} [1-\sum_{i=0}^{i=5}c_{i} (K/a)^{i}]
\]
Inserting this result into eq.(17) and using the known values of the constants $c_{i}$, one could finally get the full expression for the probability of finding a aprticle within  a given length interval in a $1D$ finite potential well. Changes of this probaility with external pressure can (in principle) be determined by calculating the derivative $dR/dP$. Such an analysis would be simplified in various limiting cases. For example, assuming that $\beta$ is small (which physically corresponds to a particle whose energy is near the top of the well), it can be shown that
\begin{equation}
R=\gamma[1+\frac{1}{3}(a\beta)^{2}(\gamma^{2}-1)]
\end{equation}
At the end of this letter,it can be stated that its purpose has not been to present completely new calculations,but to illustrate, on a well known example, an important physical concept: the influence of the external pressure on the enery levels of microscopic systems. Astrophysical consequences of this influence have already been discussed (such as [5] and references given there).


}


{









{}
\end{document}